# Edge State, Band Topology, and Time Boundary Effect in the Fine-Grained Categorization of Chern Insulators

H. C. Wu,[1,2] H. S. Xu,[1] L. C. Xie,[1] and L. Jin[1,*]

[1]School of Physics, Nankai University, Tianjin 300071, China
[2]School of Physics, Zhengzhou University, Zhengzhou 450001, China

We predict novel topological phases with broken time-reversal symmetry supporting the coexistence of opposite chiral edge states, which are fundamentally different from the photonic spin-Hall, valley-Hall, and higher-order topological phases. We find a fine-grained categorization of Chern insulators, their band topologies characterized by identical Chern number are completely different. Furthermore, we prove that different topologies cause zeros in their Bloch wavefunction overlaps, which imprint the band gap closing and appear at the degenerate points of topological phase transition. The Bloch wavefunction overlaps predict the reflection and refraction at a topological time boundary, and the overlap zeros ensure the existence of vanishing revival amplitude at critical times even though different topologies before and after the time boundary have identical Chern number. Our findings create new opportunities for topological metamaterials, uncover the topological feature hidden in the time boundary effect as a probe of topology, and open a venue for the exploration of the rich physics originating from the long-range couplings.

Topological phases of matter have nontrivial band structures and support topological states robust against disorder at the edges/interfaces of photonic topological systems [1–13]. Topological invariant characterizes the topology of Bloch band and predicts the number of topological edge/interface states, known as the bulk-boundary correspondence [14–17]. This is a fundamental principle and explains why the detection of topological invariant is important in topological physics [18–23]. For two-dimensional topological phases, the Chern number as a topological invariant predicts the number of chiral edge states in the photonic topological insulators created from breaking the time-reversal symmetry [24–31]. Furthermore, the time-reversal symmetric insulating phases may have nontrivial topology although they have a zero Chern number. For example, the photonic spin-Hall phase created from breaking the time-reversal symmetry of each individual pseudo-spin is characterized by the opposite nonzero spin-Chern numbers [32–43], which predict the numbers of chiral edge states associated with the pseudo-spin-up and pseudo-spin-down. The valley is another degree of freedom similar to the spin, but the photonic valley-Hall phase is created from breaking the inversion symmetry [44–52]. In addition, the time-reversal symmetric quadrupole topological phase also has a zero Chern number [53–62]. This nontrivial higher-order topology supports the corner states. Clearly, the time-reversal symmetry preserving topological insulating phases have completely different topologies. A question naturally raises: whether there exist novel time-reversal symmetry breaking topological insulating phases with unveiled band topologies? If yes, how to characterize and probe such topological phases?

In this Letter, we predict novel Chern insulators simultaneously holding the opposite chiral edge states with broken time-reversal symmetry, which fundamentally differs from the photonic spin-Hall, valley-Hall, and higher-order topological phases. We find a fine-grained categorization of Chern insulators. The positive and negative charges of Berry connection singularities (BCSs) distinguish the topological phases and predict the numbers of clockwise and counterclockwise chiral edge states, respectively. We further find that different topologies cause zeros in their Bloch wavefunction overlaps, which predict the time reflection and refraction at a photonic time boundary. We propose the time boundary effect as a probe of topology. The time boundary effect is the temporal analogy of the spatial boundary effect based on the space-time duality. The vanishing of time reflection/refraction imprints the band gap closing in the topological phase transitions and ensures the existence of vanishing revival amplitude at critical times when the system before and after the time boundary having different topologies even through their Chern numbers are identical. The coexistence of opposite chiral edge states with different velocities create new opportunities for the development of robust photonic devices. The discovery of the fine-grained categorization of Chern insulators provides profound insight into the bulk-boundary correspondence for the novel two-dimensional topological phases of matter. The time boundary effect offers a promising platform for the detection of topology.

We consider a two-band Bloch Hamiltonian $H(\mathbf{k}) = \mathbf{B}(\mathbf{k}) \cdot \sigma + B_0(\mathbf{k})\sigma_0$, where $\mathbf{B}(\mathbf{k}) = [h_x(\mathbf{k}), h_y(\mathbf{k}), h_z(\mathbf{k})]$ is an effective magnetic field, $\mathbf{k}$ is the momentum, $\sigma = (\sigma_x, \sigma_y, \sigma_z)$ is the Pauli matrix of spin-1/2, and $\sigma_0$ is the identify matrix. $B_0(\mathbf{k})\sigma_0$ does not affect the band topology of $H(\mathbf{k})$ in the insulating phase; but topological states in the metallic phase disappear in the projection spectra where the two bands are inseparable. Topological properties of $H(\mathbf{k})$ are encoded in $\mathbf{B}(\mathbf{k})$, which may form closed surfaces in three-dimension. Any close surface wrapping the origin positively or negatively contributes to the Chern number. We find the coexistence of positive and negative wrappings, which creates the time-reversal symmetry breaking topological insulating phases that simultaneously holding the opposite chiral edge states. The Chern number is insufficient to characterize these novel band topologies, but the BCS distinguishes these topological insulating phases (Fig. 1).

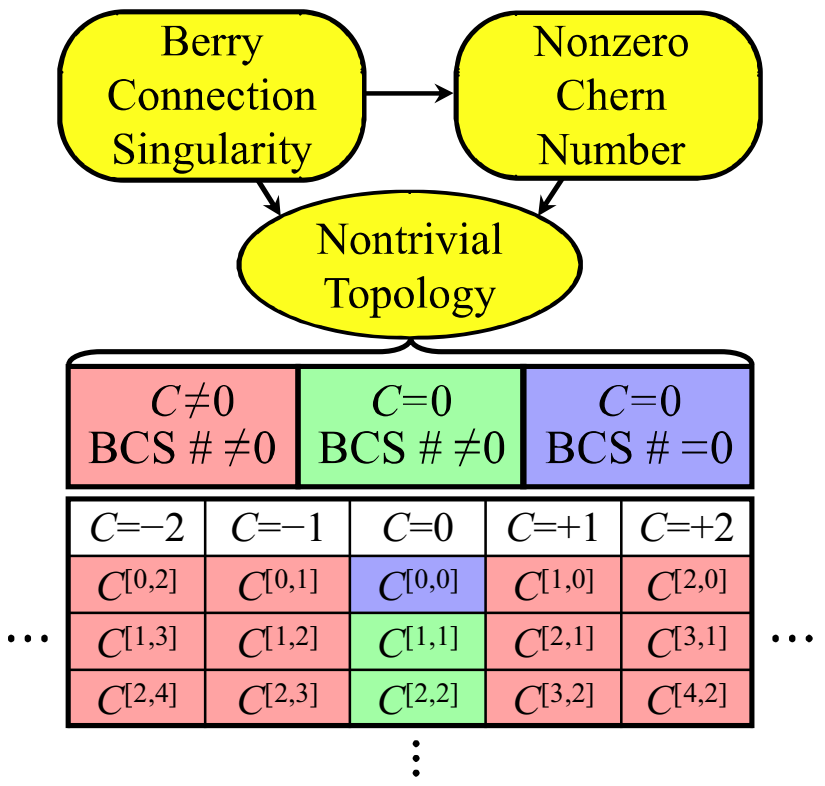

FIG. 1. Topological phases characterized by the BCSs. The BCS and nonzero Chern number cause nontrivial topology. The nonzero Chern number is attributed to the BCS. The presence of BCS may cause nontrivial topology with a nonzero Chern number (in red) or a zero Chern number (in green). The absence of BCS can have nontrivial topology (in purple). See Supplemental Material A for examples [63].

The Chern number $C_\pm = (2\pi)^{-1}\iint_{\mathrm{BZ}} \mathrm{d}^2\mathbf{k}\cdot\boldsymbol{\Omega}_\pm(\mathbf{k})$ is the integral of Berry curvature $\boldsymbol{\Omega}_\pm(\mathbf{k}) = \nabla_{\mathbf{k}} \times \mathbf{A}_\pm(\mathbf{k})$ over the entire Brillouin zone (BZ), where $\mathbf{A}_\pm(\mathbf{k}) = -i\langle\psi_\pm(\mathbf{k})|\nabla_{\mathbf{k}}|\psi_\pm(\mathbf{k})\rangle$ is the Berry connection [1]. The subscript + (−) is for the upper (lower) band. The Bloch wavefunction is

$$|\psi_\pm(\mathbf{k})\rangle = \frac{1}{\sqrt{2h(\mathbf{k})[h(\mathbf{k})\pm h_z(\mathbf{k})]}}\begin{pmatrix} h_z(\mathbf{k})\pm h(\mathbf{k}) \\ h_x(\mathbf{k})+ih_y(\mathbf{k})\end{pmatrix}, \tag{1}$$

for the band energy $\varepsilon_\pm(\mathbf{k}) = B_0(\mathbf{k}) \pm [h_x^2(\mathbf{k}) + h_y^2(\mathbf{k}) + h_z^2(\mathbf{k})]^{1/2} = B_0(\mathbf{k}) \pm h(\mathbf{k})$. The Berry connection $\mathbf{A}_\pm(\mathbf{k})$ has the non-analytic points (BCSs) at

$$[h_x(\mathbf{k}), h_y(\mathbf{k}), h_z(\mathbf{k})] = [0, 0, \mp h(\mathbf{k})]. \tag{2}$$

The BCS on the band inversion surface $h_z(\mathbf{k}) = 0$ is the degenerate point $h(\mathbf{k}) = 0$ [64]. When the BCS moving across the band inversion surface, the band gap closes and reopens associated with the topological phase transition [2, 65]. The charge of BCS $c_s = (2\pi)^{-1}\oint_{l_s}[h_y(\mathbf{k})dh_x(\mathbf{k}) - h_x(\mathbf{k})dh_y(\mathbf{k})]/[h_x^2(\mathbf{k})+h_y^2(\mathbf{k})]$ is associated with the charge of degenerate point, being a positive charge +1 or a negative charge −1. The closed loop $l_s$ in the BZ encloses only one BCS.

The singularities of $\mathbf{A}_+(\mathbf{k})$ and $\mathbf{A}_-(\mathbf{k})$ appear at $h_z(\mathbf{k}) = -h(\mathbf{k}) < 0$ and $h_z(\mathbf{k}) = +h(\mathbf{k}) > 0$, respectively. The regions $h_z(\mathbf{k}) < 0$ in the BZ are denoted as $D_+$ in yellow for the upper band and the regions $h_z(\mathbf{k}) > 0$ in the BZ are denoted as $D_-$ in white for the lower band.

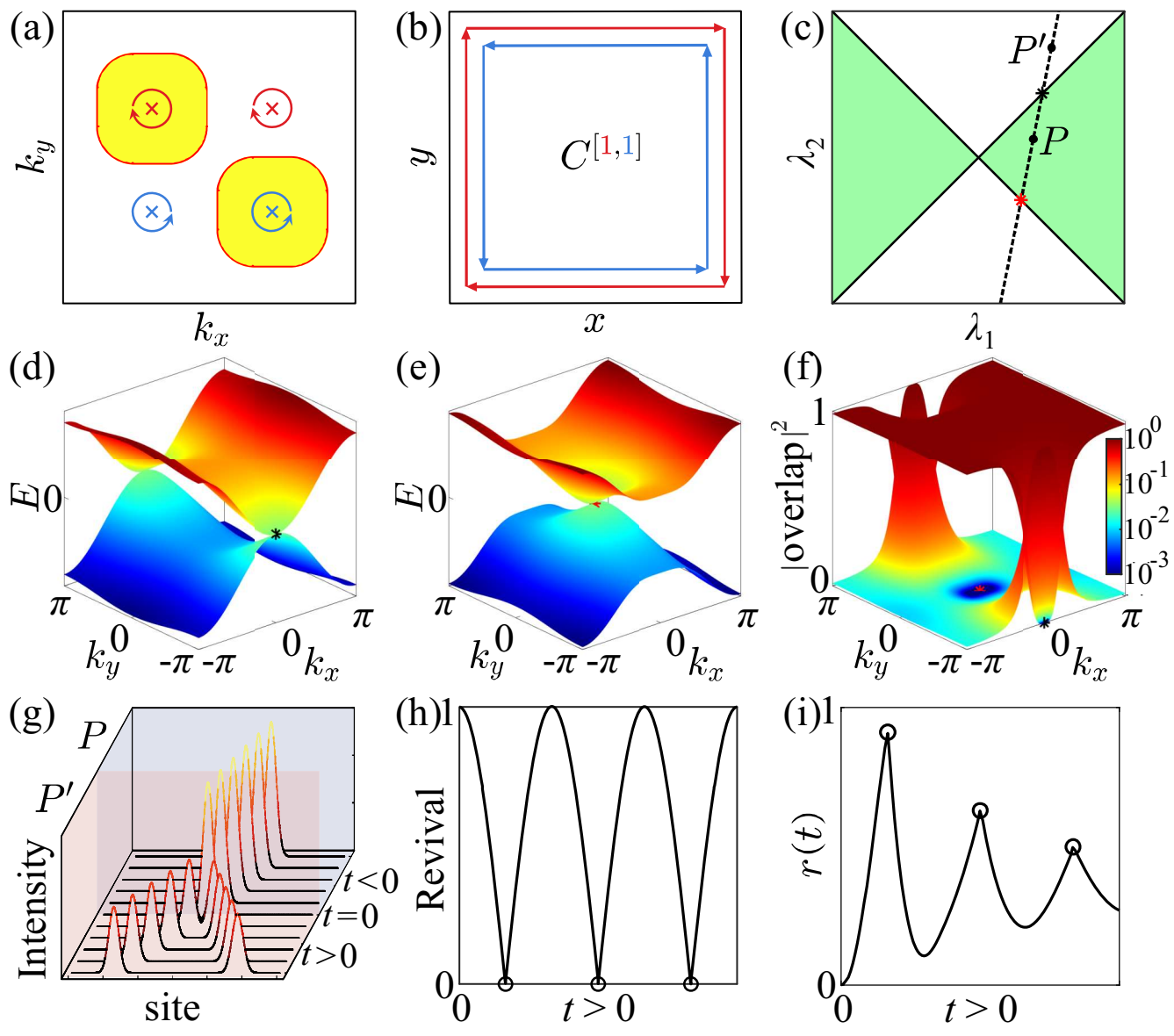

FIG. 2. Schematics of the edge state, band topology, and time boundary effect. (a) BCSs in the BZ. The red curves are the band inversion surface. (b) Coexistence of opposite chiral edge states. (c) Phase diagram in the parameter space $\lambda_1$-$\lambda_2$. The black solid lines indicate the topological phase transition. (d) [(e)] Energy bands at the black (red) star in (c). (f) Wavefunction overlaps between identical (different) Bloch bands for the system at the points $P$ and $P'$ are mainly in red (blue). (g) Equal time reflection and refraction for the initial state at a critical momentum. (h) Revival amplitude. (i) Rate function. In (g)-(i), the parameters before and after the time boundary are $P$ and $P'$.

The sum of all the singularity charges in the area $D_+$ ($D_-$) is the upper (lower) Chern number $C_+^{[m,n]} = m - n$ ($C_-^{[n,m]} = n - m$) [66]. We use the upper band Chern number to describe the topological phase and remove the band index for conciseness. The superscripts in $C^{[m,n]}$ represent $m$ positive and $n$ negative BCSs, being the minimum numbers of singularities under any chosen gauge, is equal to the numbers of positive and negative wrappings in $\mathbf{B}(\mathbf{k})$. Figure 2(a) shows the BCSs in the BZ for the phase $C^{[1,1]}$. The red (cyan) cross in the yellow area indicates the singularity with positive (negative) charge +1 (−1) and predicts a clockwise (counterclockwise) propagating chiral edge state [Fig. 2(b)].

The Chern number is zero if the Berry connection is smooth. The presence of BCSs causes nontrivial topology. The BCSs originate from the degenerate points. The change on the total positive/negative charge of BCSs before and after the topological phase transition is the total positive/negative charge of degenerate points at the topological phase transition. The nonzero number of BCSs predicts the existence of gapless edge states. At the interface between two topological areas, the number of clockwise/counterclockwise propagating edge states is equal to the difference between their total positive/negative charges of BCSs. The BCSs present in the novel Chern insulators for $mn \neq 0$ and the conventional Chern insulators for $mn = 0$ (i.e., $C^{[m,0]}$ or $C^{[0,n]}$). By contrast, the two-dimensional Zak phase is valid when the entire BZ does not have any BCS in the phase $C^{[0,0]}$. The nonzero two-dimensional Zak phase predicts the existence of in-gap edge states [53, 54]. These arguments are also valid for the multi-band Chern insulators.

Topological phases $C^{[m,n]}$ and $C^{[m',n']}$ with different positive and/or negative charges have different band topologies, which lead to zeros in their wavefunction overlaps $\langle \psi'_{q'}(\mathbf{k})|\psi_q(\mathbf{k})\rangle$ (the subscripts $q, q'$ are the band indices). The number of overlap zeros in the BZ is at least $|m-m'|+|n-n'|$, being the *minimum* number of degenerate points experienced in the topological phase transition for the band topology altering from $C^{[m,n]}$ to $C^{[m',n']}$ and vise versa. The overlap zeros present in any pair of wavefunction $|\psi_q(\mathbf{k})\rangle$ chosen from $C^{[m,n]}$ and wavefunction $|\psi'_{q'}(\mathbf{k})\rangle$ chosen from $C^{[m',n']}$; their existence is topologically-protected. Figure 2(c) shows a phase diagram. The dashed straight line along the point $P$ chosen from $C^{[m,n]}$ and the point $P'$ chosen from $C^{[m',n']}$ crosses the solid lines at the black and red stars. Then, the wavefunctions at the points $P$ and $P'$ have zeros in their overlap $\langle \psi'_{q'}(\mathbf{k})|\psi_q(\mathbf{k})\rangle$ between identical (different) bands for $q = q'$ ($q \neq q'$); and the overlap zeros appear at the degenerate points for the system at the topological phase transition as marked by the black (red) star inside (outside) the points $P$ and $P'$ (see Supplemental Material B for the proof [63]), i.e., the overlap zeros imprint the band gap closing. Figure 2(d) [Figure 2(e)] illustrates the energy bands at the black (red) star in Fig. 2(c), where the degenerate point is marked by the black (red) star. Figure 2(f) illustrates the overlaps between wavefunctions at the points $P$ and $P'$, their zeros appear at these degenerate points.

The intriguing features present at the topological time boundary $t = 0$ created via abruptly altering the system into a different topological phase. The revival amplitude $g(\mathbf{k}, t) = \langle \Psi(0)|\Psi(t)\rangle$ between the initial state $|\Psi(0)\rangle = |\psi_-(\mathbf{k})\rangle$ before the time boundary and the evolution state $|\Psi(t)\rangle = e^{-itH'(\mathbf{k})}|\psi_-(\mathbf{k})\rangle$ driven by the Bloch Hamiltonian $H'(\mathbf{k})$ after the time boundary is

$$g(\mathbf{k}, t) = \sum_{q'=\pm} |\langle \psi'_{q'}(\mathbf{k})|\psi_-(\mathbf{k})\rangle|^2 e^{-it\varepsilon'_{q'}(\mathbf{k})}, \qquad (3)$$

where the wavefunction $|\psi'_{q'}(\mathbf{k})\rangle$ of $H'(\mathbf{k})$ has the energy $\varepsilon'_{q'}(\mathbf{k})$. The wavefunction overlaps $|\langle \psi'_+(\mathbf{k})|\psi_-(\mathbf{k})\rangle|^2$ and $|\langle \psi'_-(\mathbf{k})|\psi_-(\mathbf{k})\rangle|^2$ predict the reflection and refraction at the time boundary, where the time-refracted (time-reflected) wave has the same (opposite) propagating direction with the initial wave. The wave momentum conserves before and after the time boundary and $\sum_{q'=\pm} |\langle \psi'_{q'}(\mathbf{k})|\psi_-(\mathbf{k})\rangle|^2 = 1$ [67–69].

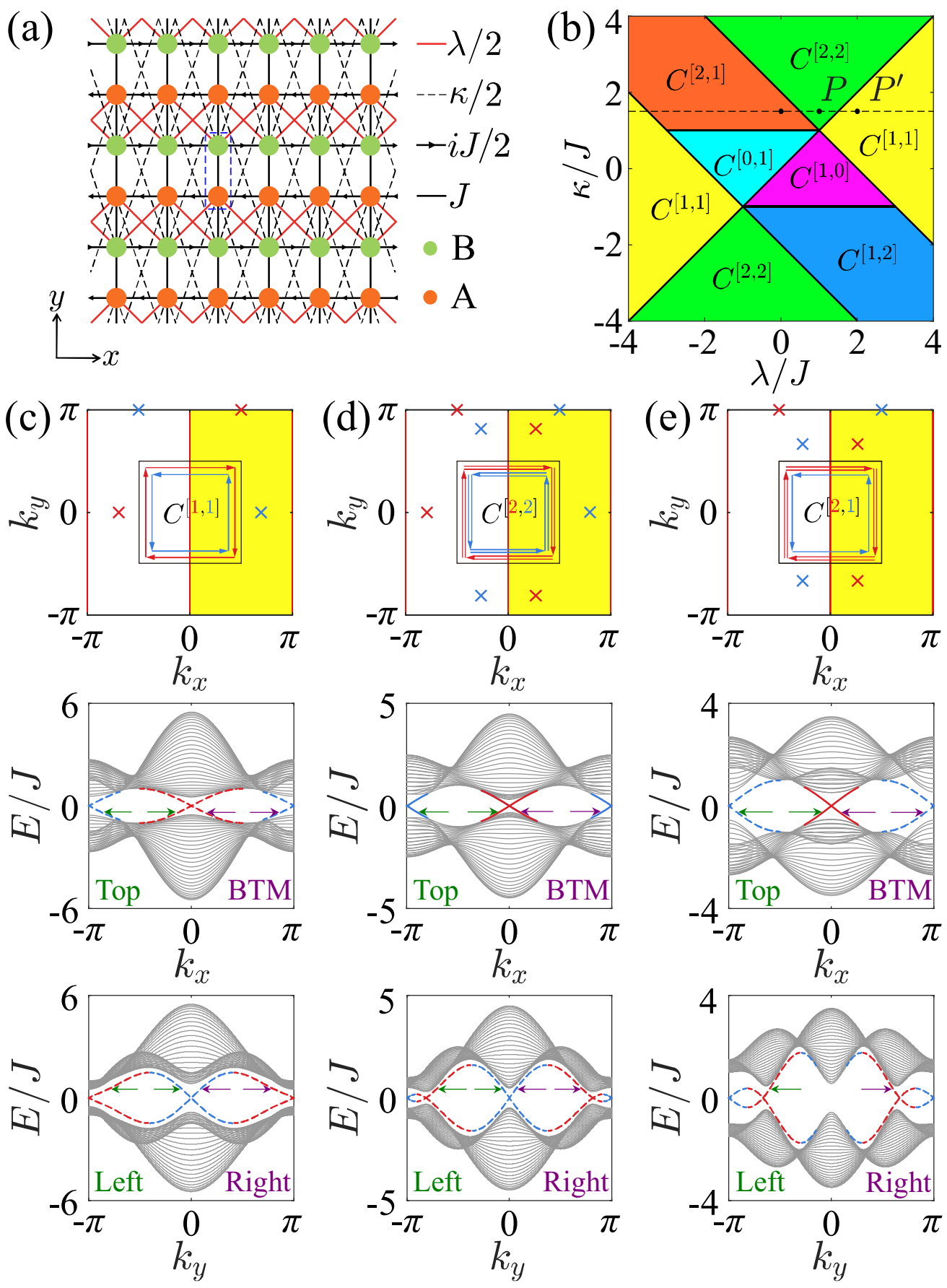

FIG. 3. (a) Square lattice. (b) Phase diagram. The distribution of BCSs, the schematic of edge states, and the projected band structures for the phases (c) $C^{[1,1]}$ at $(\lambda,\kappa) = (2J, 3J/2)$, (d) $C^{[2,2]}$ at $(\lambda,\kappa) = (J, 3J/2)$, (e) $C^{[2,1]}$ at $(\lambda,\kappa) = (0, 3J/2)$. The dashed (solid) colored lines indicate one (two-fold degenerate) edge state [70]. The abbreviation BTM stands for bottom.

Different band topologies before and after the time boundary ensure the existence of vanishing time reflection and refraction at different momenta $\mathbf{k}_1$ and $\mathbf{k}_2$. Consequently, a critical momentum $\mathbf{k}^\star = \left(k_x^\star, k_y^\star\right)$ associated with equal time reflection and refraction $|\langle \psi'_\pm(\mathbf{k}^\star)|\psi_-(\mathbf{k}^\star)\rangle|^2 = 1/2$ must exist between $\mathbf{k}_1$ and $\mathbf{k}_2$; and the number of critical momenta is at least the difference between the topological numbers before and after the time boundary $|m - m'| + |n - n'|$. Figure 2(g) illustrates the time reflection and refraction of an initial state at the critical momentum $k_c = k_y^\star$ (or $k_c = k_x^\star$) prepared in the one-dimensional projection system with $k_x = k_x^\star$ (or $k_y = k_y^\star$).

The revival amplitude for the initial state at the critical momentum periodically vanishes $g(\mathbf{k}^\star, t^\star) = e^{-itB_0(\mathbf{k})} \cos[h'(\mathbf{k}^\star)\, t^\star] = 0$ at the critical times [Fig. 2(h)]

$$t^\star = (M - 1/2)\pi/h'(\mathbf{k}^\star), \qquad (4)$$

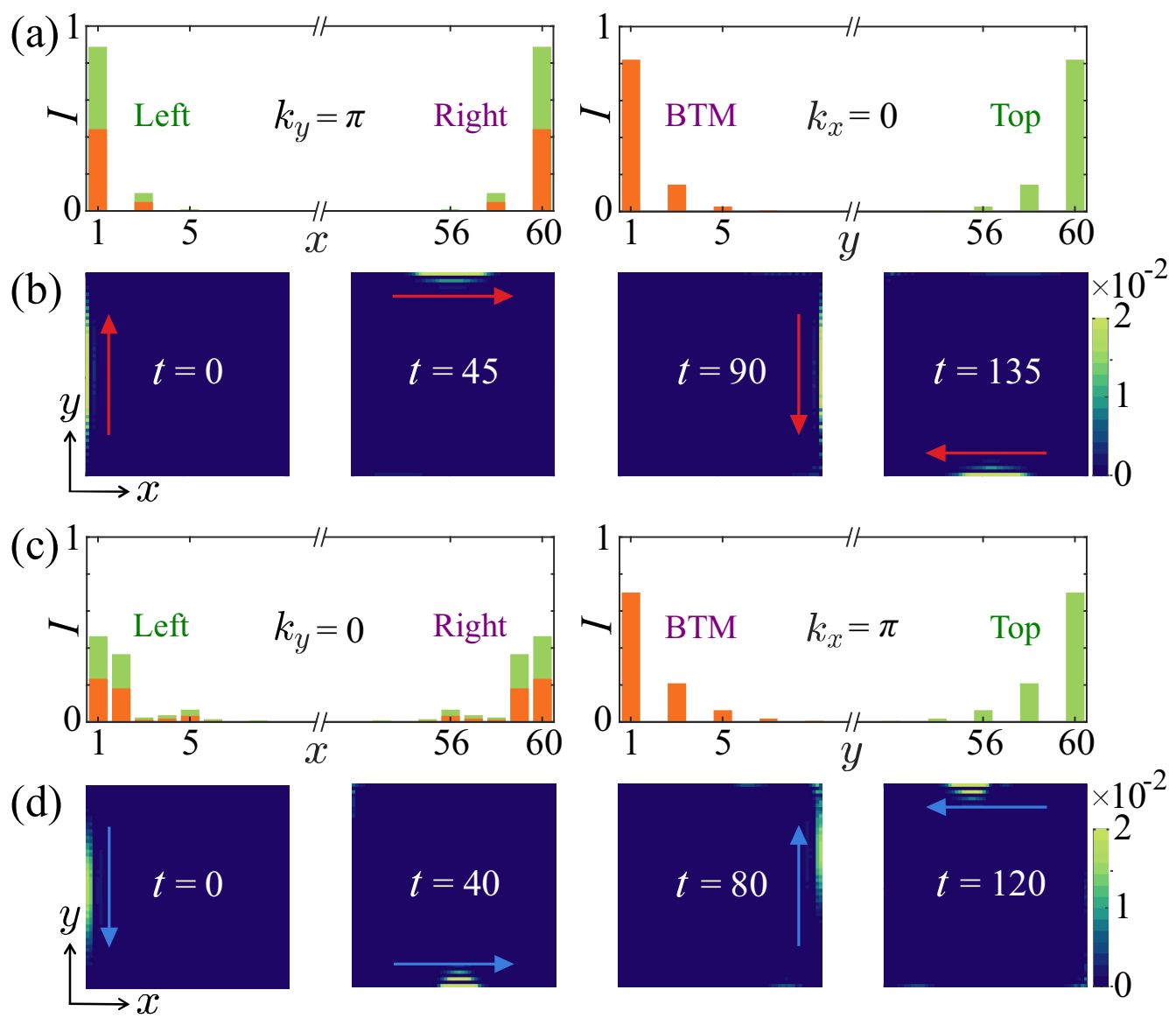

FIG. 4. The intensities of the (a) clockwise and (c) counterclockwise edge states in Fig. 3(c) at the marked momenta. The orange and green bars are for the sublattices $A$ and $B$, respectively. The propagation of left edge state excitation centered at (b) $k_y = \pi$ and (d) $k_y = 0$ in a $60 \times 60$ lattice.

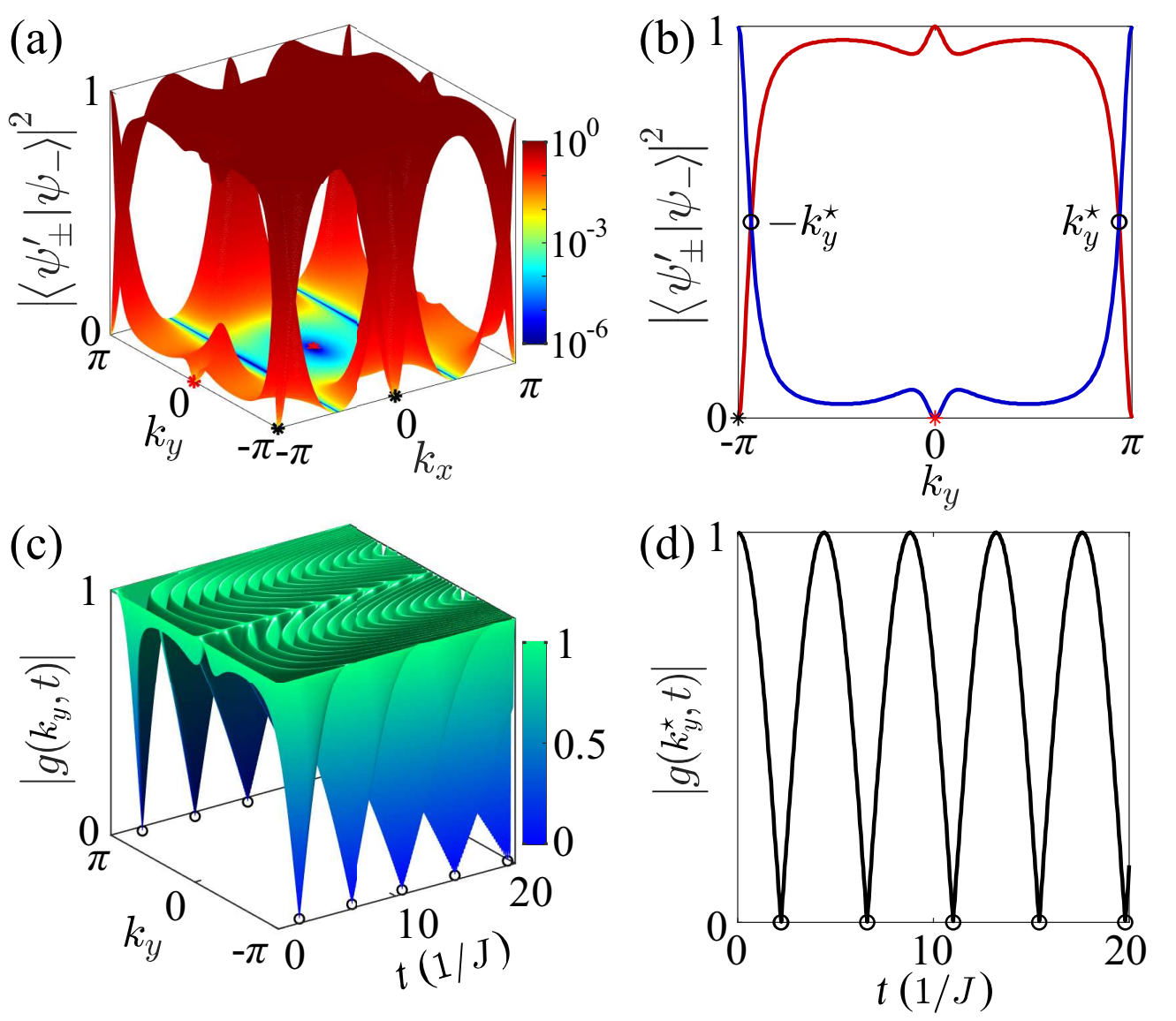

FIG. 5. (a) Overlap in the BZ. (b) Overlap at $k_x = \pi$. (c) Revival amplitude in (b) for the initial state $|\psi_-(\pi, k_y)\rangle$ before the time boundary chosen from $C^{[2,2]}$ [Fig. 3(d)] and the system after the time boundary chosen from $C^{[1,1]}$ [Fig. 3(c)]. (d) Revival amplitude in (c) at the critical momentum $k_y^\star$.

where $M \in \mathbb{N}$ is a positive integer and the revival amplitude is nonanalytic. By contrast, the revival amplitude does not vanish for the initial state not at the critical momentum. The nonanalytic rate function [Fig. 2(i)] $r(t) = -(1/N) \ln \prod_{k_y} |g(k_x^\star, k_y, t)|^2$ at the critical times for the lattice size $N$ is a dynamical analogue of the nonanalytic free energy density at the critical temperature [71–73], where the revival amplitude analogs the partition function and the rate function analogs the free energy density. The existence of nonanalytic behavior in the time boundary effect is topologically-protected.

Figure 3(a) illustrates an inversion symmetric square lattice with broken time-reversal symmetry [74–77]. The one-dimensional projection lattice after the Fourier transformation in the $x$ direction is a generalized Su-Schrieffer-Heeger model with the long-range coupling [78], which becomes the nearest-neighbor coupling if the square lattice is rearranged into bilayer [45, 60]; this model is equivalent to the coupled Su-Schrieffer-Heeger ladder [79], which is implemented in the coupled resonators [80, 81]. The one-dimensional projection lattice after the Fourier transformation in the $y$ direction is the Creutz ladder [82, 83]. The effective magnetic field in the Bloch Hamiltonian after the Fourier transformations in both directions is $h_x(\mathbf{k}) = J + (J + \lambda \cos k_x) \cos k_y + \kappa \cos k_x \cos(2k_y)$, $h_y(\mathbf{k}) = (J + \lambda \cos k_x) \sin k_y + \kappa \cos k_x \sin(2k_y)$, $h_z(\mathbf{k}) = -J \sin k_x$, and $B_0(\mathbf{k}) = 0$.

Figure 3(b) shows the phase diagram. The number of positive/negative BCSs predicts the number of clockwise/counterclockwise chiral edge states in red/cyan, counted from their crossings at the band gap closing degenerate points. The creation/annihilation of edge states is associated with the creation/annihilation of BCSs at the topological phase transition. Figures 3(c) to 3(e) exhibit the BCSs and spectra of the novel topological phases $C^{[1,1]}$, $C^{[2,2]}$, and $C^{[2,1]}$. The opposite chiral edge states in the phase $C^{[1,1]}$ and their propagations are shown in Fig. 4 (see Supplemental Material C for the detail and robustness [63]). The propagations in opposite directions have different periods. In experiments, the realization of the phase $C^{[1,1]}$ at $\kappa = 0$ only requires the next-nearest-neighbor coupling [41, 84–86]. The long-range couplings enrich the novel topological phases $C^{[m,n]}$ with $mn \neq 0$ in the photonic and acoustic metamaterials [78, 87].

Topological overlap zeros present in the Bloch wavefunctions from any two different topological phases. For example, we consider the point $P$ at $(\lambda, \kappa) = (J, 3J/2)$ chosen from $C^{[2,2]}$ and the point $P'$ at $(\lambda, \kappa) = (2J, 3J/2)$ chosen from $C^{[1,1]}$. Both topological phases have a zero Chern number and differ from $C^{[0,0]}$ [88]. In the phase diagram Fig. 3(b), the straight dashed line along the points $P$ and $P'$ crosses three solid lines of topological phase transitions. At the cross inside the points $P$ and $P'$, the degenerate points are $(k_x, k_y) = (0, \pi)$ and $(\pi, \pi)$. In Fig. 5(a), the overlap $|\langle \psi'_-(\mathbf{k}) | \psi_-(\mathbf{k}) \rangle|^2$ vanishes at these degenerate points as marked by the black stars. At the crosses outside the points $P$ and $P'$, the degenerate points are $(k_x, k_y) = (\pi, 0)$ and $(0, 0)$. In Fig. 5(a), the overlap $|\langle \psi'_+(\mathbf{k}) | \psi_-(\mathbf{k}) \rangle|^2$ vanishes at these degenerate points as marked by the red stars. In addition, two degenerate lines appear at $k_x = \pm\pi/2$ for $\lambda/J \to \pm\infty$, where $|\langle\psi'_+(\mathbf{k})\,|\psi_-(\mathbf{k})\rangle\,|^2 = 0$.

The vanishing revival amplitude in the time boundary effect witnesses the inequivalent band topologies before and after the time boundary (see Supplemental Material D for the experimental realization [63]). Figure 5(b) exhibits the reflection and refraction at the time boundary in the generalized Su-Schrieffer-Heeger model with $k_x = \pi$ for the initial state $|\psi_-(\pi, k_y)\rangle$, where the red curve is the time refraction $|\langle\psi'_-(\pi, k_y)\,|\psi_-(\pi, k_y)\rangle\,|^2$, the blue curve is the time reflection $|\langle\psi'_+(\pi, k_y)\,|\psi_-(\pi, k_y)\rangle\,|^2$, and $k_y^\star = 0.936\pi$ is the critical momentum. Figure 5(c) shows the revival amplitude $g(\pi, k_y, t)$, which periodically vanishes at the critical times $t^\star = 2.2/J$, $6.6/J$, $11.0/J$, $\cdots$ for $k_y = k_y^\star$ as indicated by the black hollow circles in Fig. 5(d). Similar behaviors exist in the time boundary effect performed in the Creutz ladder with $k_y = \pi$ for the initial state $|\psi_-(k_x^\star, \pi)\rangle$.

In summary, we proposed the novel band topologies, which enrich the kaleidoscopes of Chern insulators and gapless edge states. The coexistence of opposite chiral edge states opens a new door for the robust light transport [89]. We found that topological zeros present in the overlaps between Bloch wavefunctions from different topological phases at the degenerate points of topological phase transition and imprint the band gap closing. The overlap zeros correspond to the vanishing of time reflection/refraction [69] and ensure the existence of nonanalytic behavior in the time boundary effect as a witness of the different band topologies before and after the time boundary. The time boundaries create photonic time crystals [90], realize broadband frequency translation [91], and offer controllable light manipulation in synthetic dimension [92], our findings pave the way for exploring topology using the non-equilibrium dynamics of photons at the time boundary. It is interesting to further consider the photonic time boundary effect in non-Hermitian topological phases [93–100].

This work was supported by National Natural Science Foundation of China (Grants No. 12222504, No. 12305033, and No. 11975128).

---

* jinliang@nankai.edu.cn